\documentclass[runningheads,a4paper]{llncs}[2015/06/24]
\usepackage{cmap}
\usepackage[T1]{fontenc}
\usepackage{graphicx}
\usepackage{subfig}
\usepackage[english]{babel}
\usepackage{color}
\usepackage{algpseudocode}
\usepackage{amsmath}

\begin{document}

\title{TxProbe: Discovering Bitcoin's Network Topology Using Orphan Transactions}
%If Title is too long, use \titlerunning
\titlerunning{TxProbe: Discovering Bitcoin's Network Topology}

%Single insitute
%\author{Firstname Lastname \and Firstname Lastname}
%If there are too many authors, use \authorrunning
%\authorrunning{First Author et al.}
%\institute{Institute}

%Multiple insitutes
%Currently disabled
%
\iftrue
%Multiple institutes are typeset as follows:
\author{Sergi Delgado-Segura\inst{1, *} \and Surya Bakshi\inst{2} \and Cristina P\'erez-Sol\`a\inst{3} \and James Litton\inst{4} \and Andrew Pachulski\inst{4} \and Andrew Miller\inst{2, *} \and Bobby Bhattacharjee\inst{4}}
%If there are too many authors, use \authorrunning
\authorrunning{Sergi Delgado-Segura et al.}

\institute{
Universitat Autònoma de Barcelona\\
\and
University of Illinois Urbana-Champaign\\
\and
Universitat Rovira i Virgili\\
\and
University of Maryland\\
* Corresponding Authors: s.delgado@ucl.ac.uk, soc1024@illinois.edu
}
\fi
			
\maketitle

\begin{abstract}
  Bitcoin relies on a peer-to-peer overlay network to broadcast transactions and blocks.
  From the viewpoint of network measurement, we would like to observe this topology so we can characterize its performance, fairness and robustness.
  However, this is difficult because Bitcoin is deliberately designed to hide its topology from onlookers.
  Knowledge of the topology is not in itself a vulnerability, although it could conceivably help an attacker performing targeted eclipse attacks or to deanonymize transaction senders.
  %So far, Bitcoin developers have patched behavior as soon as it is found to be useful, frustrating measurement efforts.
 
  In this paper we present TxProbe, a novel technique for reconstructing the Bitcoin network topology.
  TxProbe makes use of peculiarities in how Bitcoin processes out of order, or ``orphaned'' transactions.
  We conducted experiments on Bitcoin testnet that suggest our technique reconstructs topology with precision and recall surpassing 90\%. We also used TxProbe to take a snapshot of the Bitcoin testnet in just a few hours. TxProbe may be useful for future measurement campaigns of Bitcoin or other cryptocurrency networks.
\end{abstract}

%\begin{keywords}
%keyword1, keyword2
%\end{keywords}

%%%%%%%%%%%%%%%%%%%%%%%%%%%%%%%%%%%%%%%%%%%%%%%%%%%%%%%%%%%%%%%%%%%%%%%%%%%%%%%
% Introduction
%%%%%%%%%%%%%%%%%%%%%%%%%%%%%%%%%%%%%%%%%%%%%%%%%%%%%%%%%%%%%%%%%%%%%%%%%%%%%%%
\section{Introduction}
\label{sec:intro}

Bitcoin builds on top of a peer-to-peer (P2P) network to relay transactions and
blocks in a decentralized manner. Broadcast is the routing scheme chosen to
propagate transactions and blocks over the network, in order to spread the
information as quick as possible and facilitate agreement on a common state. The topology of the Bitcoin network is unknown by design and it is built to mimic a random network. While knowing the topology of the network does not pose a threat by itself, it eases the performance of several network based attacks, such as eclipse attacks~\cite{heilman2015eclipse,nayak2016stubborn}, or attacks on users anonymity~\cite{koshy2014FC,biryukov2014deanonymisation}. 

On top of that, a study of the network topology may reveal to what extent the network is really decentralized, whether there exist supernodes, bridge nodes, potential points of failure, etc.

In this paper we present TxProbe, a technique to infer the topology of the
publicly reachable Bitcoin network. Nodes of the non-reachable network, such as nodes behind NAT or firewalls, or nodes not accepting incoming connections will not be inferred with our technique. Our work builds on prior work in exploiting Bitcoin network side channels as measurement techniques, but exploits a new side channel involving the handling of orphan transactions (transactions that arrive out of order). 

To validate our technique, we have conducted an experiment in which our custom
node is connected to our own ground truth nodes (running Bitcoin Core software).
We then check whether we were able to get the connections of such nodes. On top
of that, a scan of the entire live network has been performed resulting on a
snapshot of the Bitcoin testnet. Finally, a comparative analysis of the obtained
testnet graph against similar random graphs is provided to quantify whether or
not the network resembles a random network.

TxProbe is an active measurement technique, and we have not conclusively ruled out that it could interfere with ordinary transactions. We have therefore limited our measurement and validate activities to the Bitcoin testnet. The technique could be used in the future to infer the topology of Bitcoin or any alt-coin sharing its network protocol, including Bitcoin Cash, Litecoin or Dogecoin.

%%%%%%%%%%%%%%%%%%%%%%%%%%%%%%%%%%%%%%%%%%%%%%%%%%%%%%%%%%%%%%%%%%%%%%%%%%%%%%%
% Related work
%%%%%%%%%%%%%%%%%%%%%%%%%%%%%%%%%%%%%%%%%%%%%%%%%%%%%%%%%%%%%%%%%%%%%%%%%%%%%%%
\section{Related work}
\label{sec:related}

Network topology inference is a topic that has been previously analysed in
several other works. Biryukov et al.~\cite{biryukov2014deanonymisation} showed how a node could be uniquely identified by a subset of its neighbourhood, and how the neighbourhood could be easily inferred by checking the address messages propagation throughout the network. Biryukov et al.~\cite{Biryukov215BitcoinTor} also showed how using Tor to guard against the aforementioned technique was not useful, and it could even ease the deanonymization process. 

The use of address message propagation along with timestamp analysis was used by Miller et al.~\cite{miller2015coinscope} to infer the topology of the Bitcoin network. The analysis highlighted how the network did not behave as a random graph but, instead, it was filled with several influential nodes representing a disproportionate amount of mining power. Their AddressProbe technique took advantage of the two-hour penalty applied to received address messages from connected peers. However, the two-hour penalty was removed from the Bitcoin Core nodes after 0.10.1 release ~\cite{core:_bitcoin_core101,addr_patch}, reducing the fingerprint left by address messages, and therefore, making AddressProbe no longer useful to infer the topology of the network. 

Neudecker et al.~\cite{Neudecker2016Timing} performed timing analyses of the transaction propagation to infer the topology of the Bitcoin network with a substantial precision and recall ($\sim$ 40\%). 

Network information from the P2P network has also been used, alongside with address clustering heuristics, to check whether such information could be useful in the deanonymization of Bitcoin users~\cite{Neudecker2017Clustering}. The study shows how while most of the network information cannot ease the address clustering process, a small number of users show correlations that may make them vulnerable to network based deanonymization attacks.

A recent proposal by Grundmann et al.~\cite{grundmann2018topology} has shown how transaction accumulation of double-spending transactions can also be used to infer the neighbourhood of a targeted node with precision and recall as high as 95\%.

Finally, Efe Gencer et al.~\cite{gencer2018decentralization} have presented a comparative analysis of the decentralization on two of the most popular cryptocurrencies to the date, Bitcoin and Ethereum, using application layer information obtained from the Falcon Network. Their results show how around 56\% of Bitcoin nodes are run in datacenters. On top of that, their study highlights how the top four Bitcoin miners control more than the 54\% of the mining power.
%%%%%%%%%%%%%%%%%%%%%%%%%%%%%%%%%%%%%%%%%%%%%%%%%%%%%%%%%%%%%%%%%%%%%%%%%%%%%%%

%%%%%%%%%%%%%%%%%%%%%%%%%%%%%%%%%%%%%%%%%%%%%%%%%%%%%%%%%%%%%%%%%%%%%%%%%%%%%%%
% Background
%%%%%%%%%%%%%%%%%%%%%%%%%%%%%%%%%%%%%%%%%%%%%%%%%%%%%%%%%%%%%%%%%%%%%%%%%%%%%%%
\section{Background}
\label{sec:background}
\label{sec:tx_propagation}

In this section we give an overview of Bitcoin's transaction propagation behavior. Since our TxProbe technique relies on subtleties of this process, we go into detail on just the relevant parts.
	
\subsection{Three-round transaction propagation}
\label{sec:net_messages}
\begin{figure}[tb]
	\centering	
	\includegraphics[scale=0.4]{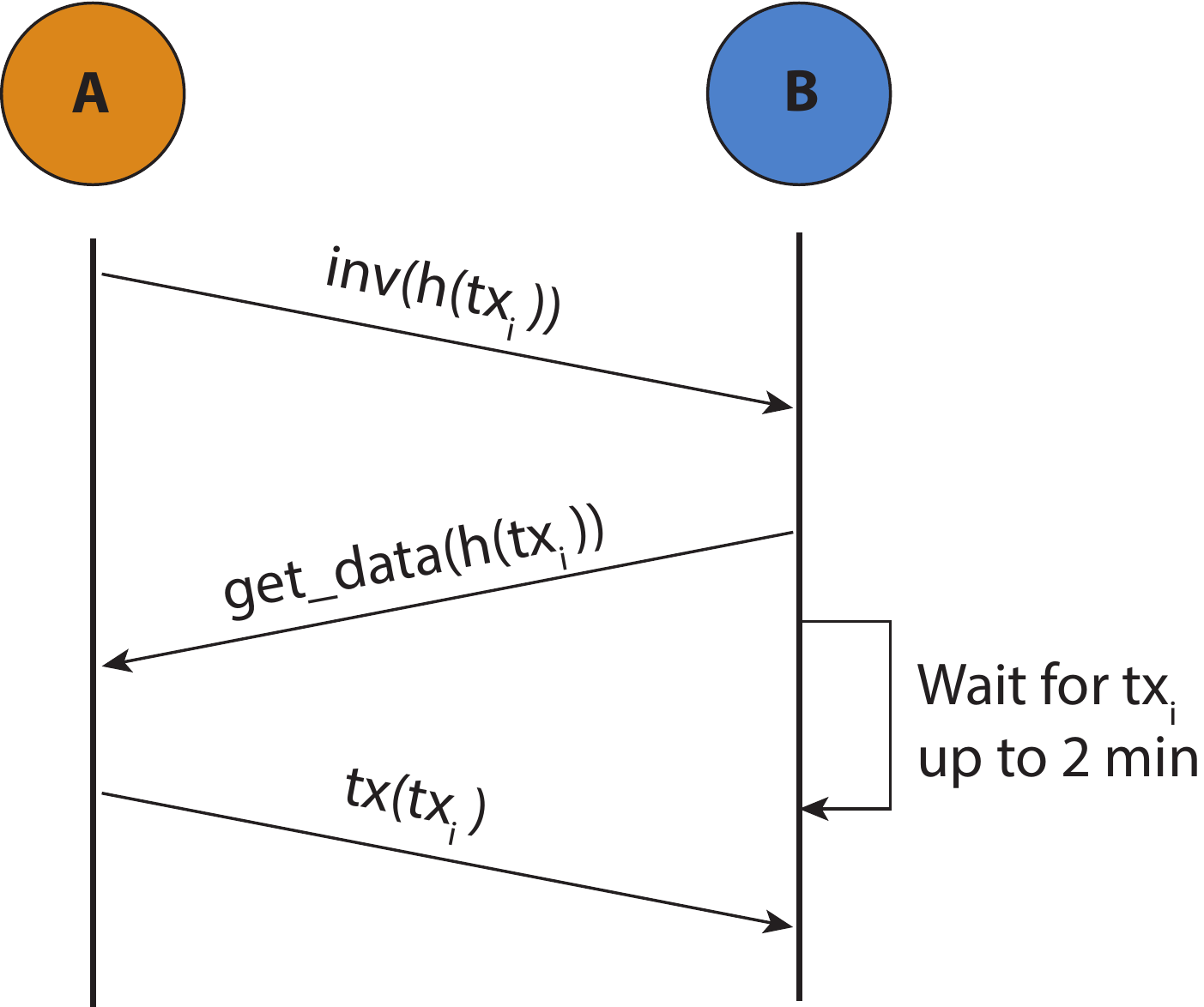}
	\caption{Three-step protocol used to forward transactions in Bitcoin.}
	\label{fig:three_step_prot}
\end{figure}

Bitcoin nodes propagate transactions by flooding, such that each node relays data about each transaction to every one of its peers.
However, to minimize network traffic, nodes follow a three-step protocol, first sending just the transaction hash (32 bytes) and only sending the entire transaction (range from a few hundred bytes up to tens of kilobytes) if it is requested. This protocol is depicted in Figure \ref{fig:three_step_prot}. In more detail, the three steps are:

\begin{itemize}
	\item 	\textbf{Inventory messages (inv)} are used to announce the
          knowledge of one or more transactions or blocks. When a node receives
          (or generates) a new transaction or block he announces it to his peers
          by creating an \texttt{inv} message containing the transaction
          hash. Those peers who do not know about the announced item will ask
          for it back using a \texttt{getdata} message. Furthermore, when a node receives an \texttt{inv} message asking for a certain item, and he requests it back using a \texttt{getdata} message, the requester will wait up to 2 minutes for the node offering the item to respond back with it. Any other request offering the same item will be queued and only responded, first in first out, if the first node fails to reply.
	\item \textbf{Get data messages (getdata)} are used by Bitcoin nodes to request transactions and blocks to their peers. Such messages are sent as a response to the aforementioned \texttt{inv} messages when the receiver of the latter is interested in any of the offered items.
	\item \textbf{Transaction messages (tx)} are used to send transactions between peers. They are usually sent as a response to a \texttt{getdata} message. In contrast to the previously introduced messages, \texttt{tx} messages always contain a single transaction.
\end{itemize}

\subsection{Mempool and Unspent Transaction Outputs (UTXOs)}
A Bitcoin node validates each transaction it receives before relaying to its peers as described above.
A valid transaction must have correct signatures, and must only spend existing and currently-unspent coins. Otherwise the transaction is discarded and not propagated further.

To aid in validation, each Bitcoin full node maintains a view of the current set of available coins (the \texttt{utxo set}).
It also maintains a collection of pending transactions, called \texttt{mempool}, all of which  have been validated against the \texttt{utxo set} and contain no double-spends amongst themselves.

Much of the complexity in the Bitcoin software, and the behavior we exploit in TxProbe, involves handling special cases during validation.
Hence when a transaction is received, it is validated against the current \texttt{utxo set}.
Since mempool is also kept free of double-spends, when a Bitcoin node receives a second transaction that spends the same coin as a transaction held in \texttt{mempool}, the second transaction is simply discarded.%
\footnote{There is a special case, called replace-by-fee (RBF)~\cite{BIP125}, in which a double-spending transaction replaces a previous transaction as long as the previous transaction is flagged to allow this and if the new transaction pays a larger fee. This does not affect the TxProbe technique.}

\subsection{Handling orphan transactions}
Sometimes a node receives  transactions out of order.
A transaction is considered an ``orphan'' if it is received prior to its direct ancestors, i.e. it spends a coin that is not yet part of the blockchain or in \texttt{mempool}.
Since orphan transactions cannot be validated until the parent arrives, they are not immediately relayed to peers.
Instead, orphan transactions are stored in a buffer, \texttt{MapOrphanTransactions} so that when the parent arrives it can be validated without re-requesting it from the network.

To point out a detail relevant to our TxProbe technique:
If a node receives a notice about a transaction from a peer (an \texttt{inv} message), but that transaction has already been stored as an orphan, then that transaction will be omitted from subsequent \texttt{get\_data} messages.
Looking ahead, this behavior enables our measurement node to probe whether an orphan transaction has already been received or not.
 We discuss other details about \texttt{MapOrphanTransactions}, such as eviction policies, later on when discussing optimizations to TxProbe.

%%%%%%%%%%%%%%%%%%%%%%%%%%%%%%%%%%%%%%%%%%%%%%%%%%%%%%%%%%%%%%%%%%%%%%%%%%%%%%%

%%%%%%%%%%%%%%%%%%%%%%%%%%%%%%%%%%%%%%%%%%%%%%%%%%%%%%%%%%%%%%%%%%%%%%%%%%%%%%%
% TxProbe
%%%%%%%%%%%%%%%%%%%%%%%%%%%%%%%%%%%%%%%%%%%%%%%%%%%%%%%%%%%%%%%%%%%%%%%%%%%%%%%
\section{Inferring the Bitcoin network topology}
\label{sec:txprobe}

In this section we explain our technique for inferring the topology of Bitcoin's reachable peer-to-peer network, making use of the subtleties of transaction propagation in Bitcoin as described earlier, and in particular conflicting transactions and orphan transactions.
We start by introducing a basic edge inference technique that tests for an edge between a single pair of peers. Later we discuss how to scale the technique up to take network-wide snapshots efficiently.

\begin{figure}[tb]
	\centering
	\subfloat[Basic positive edge inferring technique between two nodes.\label{fig:basic_pos_infer}]{{\includegraphics[scale=.45]{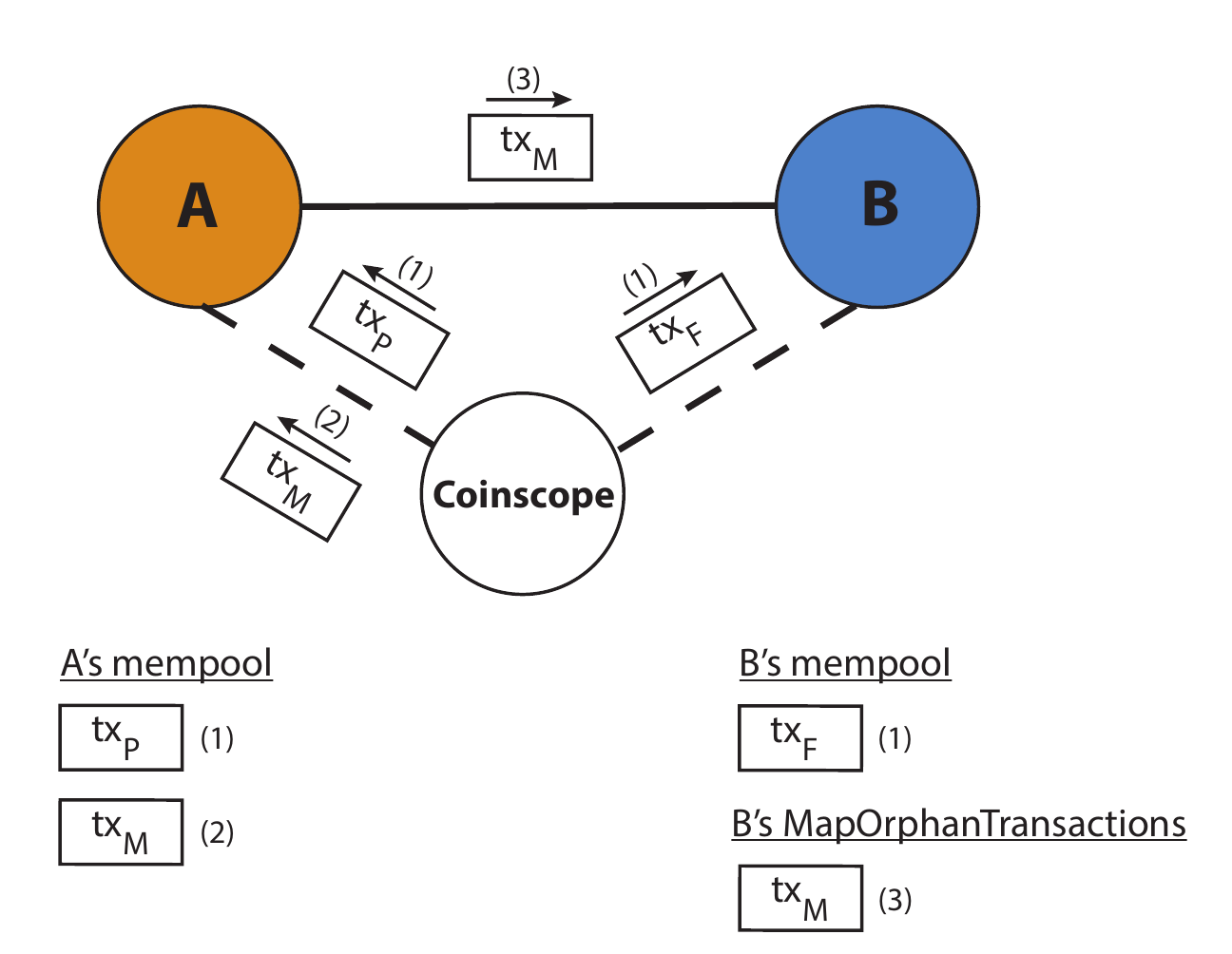}}}%
	\hfill
	\subfloat[Basic negative edge inferring technique between two nodes.\label{fig:basic_neg_infer}]{{\includegraphics[scale=.45]{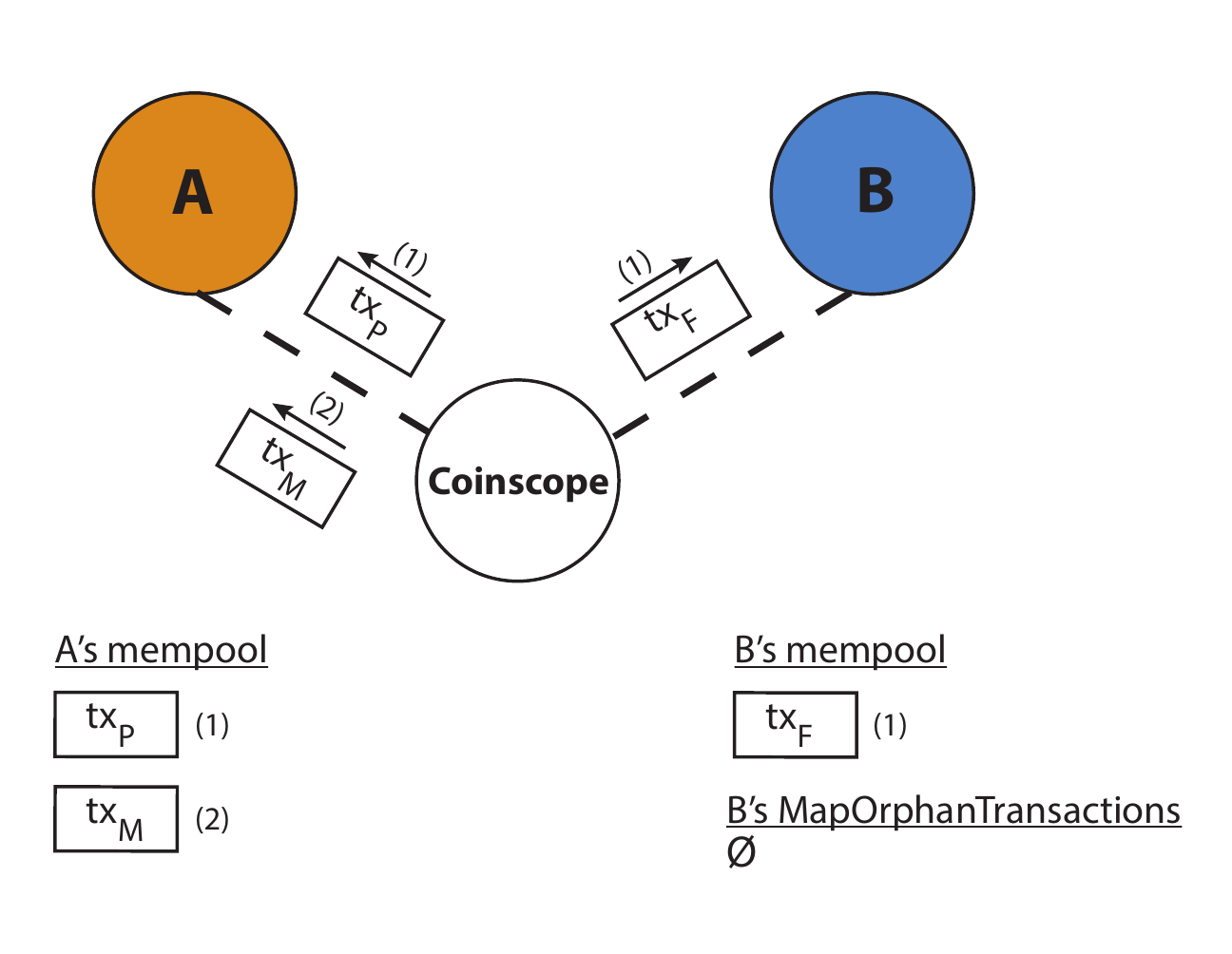}}}%
	\caption{Basic edge inferring technique.}
\end{figure}

\subsection{Basic edge inferring technique}
\label{sec:basic_protocol}
To explain the main idea behind our technique, we start by describing a scenario
in which our Coinscope measurement node is connected to two nodes $A$ and $B$,
and we want to check if there exists an edge between them. Note that such a scenario is not realistic, but we will discuss real cases later on. First we create a
pair of double spending transactions referred to as the parent ($tx_P$) and the flood ($tx_F$). We send $tx_P$ to $A$ and $tx_F$ to $B$ and assume that both transactions arrive to their destination at the same time, so $A$ will reject $tx_F$ if $B$ sends it to him and vice versa. Now we create a third transaction, the marker ($tx_M$), that spends from $tx_P$ and we send it to $A$. On receiving $tx_M$, $A$ will forward it to all his peers. If the edge between the two nodes exists, as depicted in Figure~\ref{fig:basic_pos_infer}, $B$ will receive the transaction. It is worth noting that $B$ does not know about $tx_P$, so $tx_M$ will be flagged as orphan and not relayed any further. On the contrary, if the edge between nodes $A$ and $B$ does not exist, as depicted in Figure~\ref{fig:basic_neg_infer}, $tx_M$ will never be sent to $B$. 

At this point we can check if the connection between the two nodes exists.  To do so, we ask $B$ about $tx_M$ by sending him an \texttt{inv} message containing $tx_M$'s hash. If the connection between the two nodes exists, $B$ will have $tx_M$ stored in his \texttt{MapOrphanTransactions} pool, so he will not request it back. On the other hand, if the edge does not exist, $B$ will respond with a \texttt{getdata} message containing $tx_M$'s hash.

\begin{figure}
	\centering
	\includegraphics[scale=0.5]{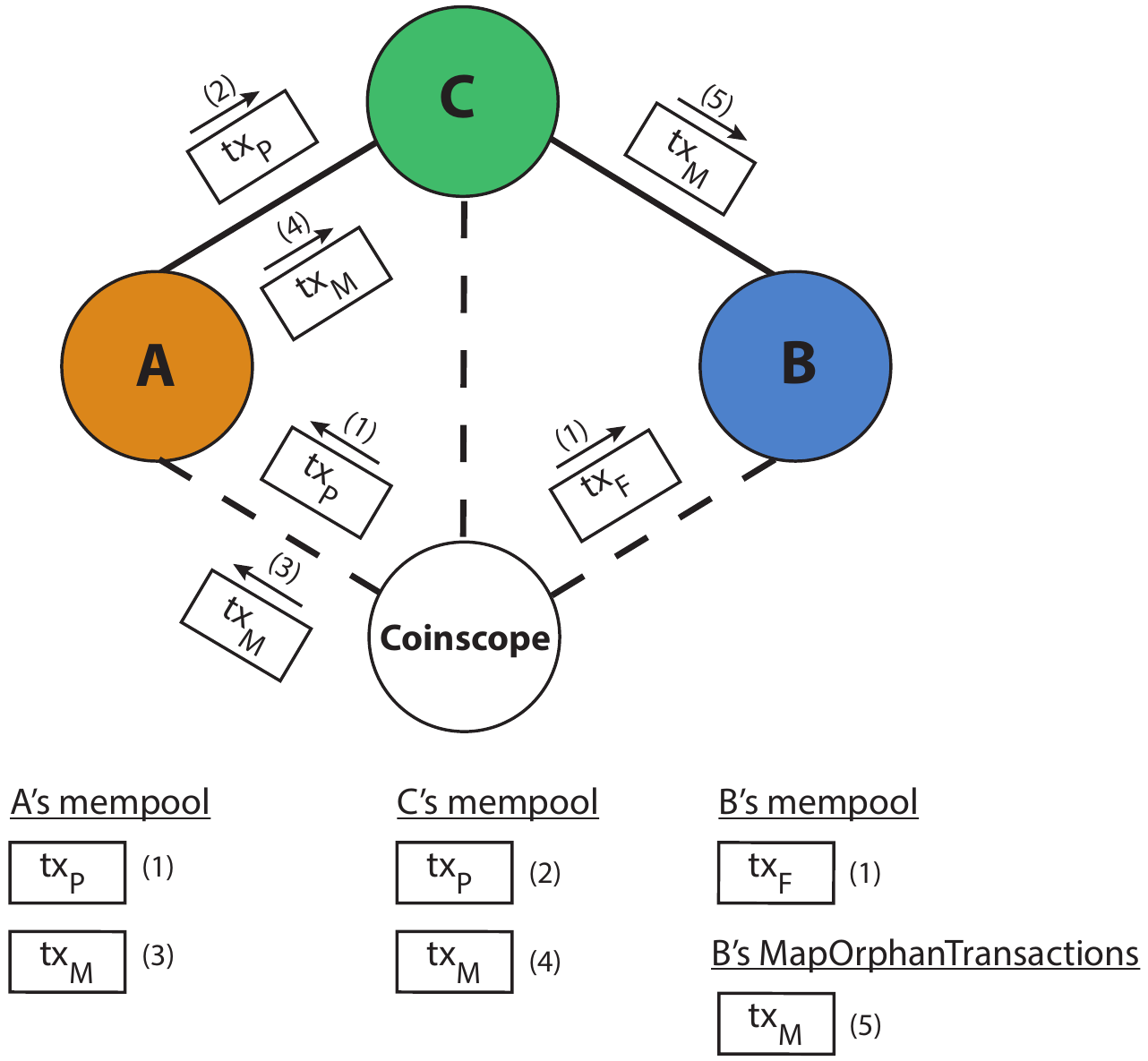}
	\caption{Incorrect edge inferring with three nodes.}
	\label{fig:infer_wrong}
\end{figure}

While this basic technique works in the most simple scenario, namely with two nodes potentially connected only between them, it can drastically fail if just one additional node is added to the picture. Let's see what happens if we connect one additional node $C$, as depicted in Figure~\ref{fig:infer_wrong}, and repeat the procedure. Since $A$ is connected to $C$, $tx_P$ and $tx_M$ will be forwarded to him, $C$ will treat both transactions as regular ones, and forward them to $B$, who will reject $tx_P$ as double spending of $tx_F$, but accept $tx_M$ as orphan. Ultimately, we will ask $B$ about $tx_M$, and infer a non-existing edge between him and $A$. 

Such a basic example highlights the first main issue of the basic approach: \textbf{isolation}. We need to ensure that $tx_P$ remains only in the node we have sent it to. Otherwise, we may end up inferring non-existent edges. Moreover, this basic technique also builds on top of another fragile property: \textbf{synchronicity}. If node $A$ receives $tx_P$ before $B$ receives $tx_F$, $A$ can forward $tx_P$ to $B$, and the latter will end up rejecting $tx_F$ upon reception, making the technique to fail. Finally, the basic technique lacks \textbf{scalability}. Assuming we can sort out the two previous issues, for every three transactions created we will be able to infer at most the whole neighbourhood of a single node. Inferring the topology of the whole reachable network will require creating almost three transactions per node, namely $3\cdot(n-1)$ where $n$ is the number of reachable nodes in the network.
%$3*\mathit{len}(\mathit{reachable\_network}) -1$. 
In order to solve the three aforementioned problems, we have created a technique called TxProbe.

\subsection{TxProbe}

TxProbe is a topology inference technique that uses double spending and orphan transactions to check the existence of edges between a pair of nodes. TxProbe can be used to infer the topology of several cryptocurrency P2P networks, as long as they share the network protocol and orphan transactions handling with Bitcoin (i.e. Bitcoin Cash, Litecoin, ZCash, etc). In contrast to recently proposed techniques, such as \cite{grundmann2018topology}, TxProbe is intended to perform full network topology inference, instead of targeted neighbourhood discovery, even though the latter can be also achieved. TxProbe builds on the aforementioned basic edge inferring technique solving its three main downsides:

Regarding \textbf{isolation} and \textbf{synchronicity}, TxProbe uses Coinscope, the observation and testing framework introduced by Miller et al. in \cite{miller2015coinscope}, to maintain connections with all reachable nodes and performs the \texttt{invblock} technique (proposed also in \cite{miller2015coinscope}) to ensure that a target transaction will remain in a target node. With regard to \textbf{scalability}, TxProbe takes advantage of the \texttt{MapOrphanTransactions} pool management to perform multiple nodes neighbourhood discovery at the same time.

We now describe the main components of the TxProbe technique.
In a single trial, we break the network nodes into two groups, the \texttt{source set} and the \texttt{sink set}, where we aim to infer all the connections between \texttt{source set} nodes and \texttt{sink set} nodes. The \texttt{source set} will be usually smaller than the \texttt{sink set}, and should at least be less than the size of the \texttt{mapOrphanTransactions} pool.

\subsection*{Setup}

\begin{figure}[tb]
	\centering
	\includegraphics[scale=0.3]{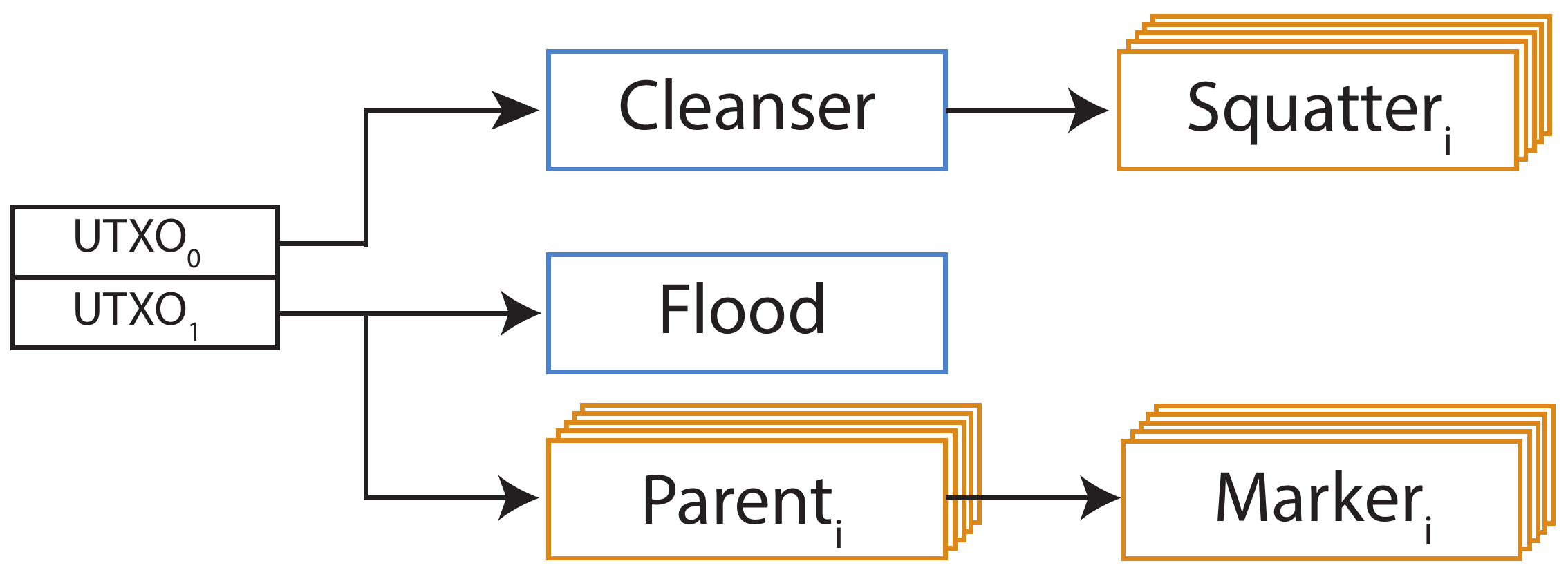}
	\caption{High level representation of the transactions created in TxProbe}
	\label{fig:tx_scheme_all}
\end{figure}

\paragraph{\textbf{Create conflicting transactions:}} First, we need to create the set of conflicting transactions, namely the parents, markers, and flood transactions. This time we are not targeting a single node to infer his peers (as we did with $A$ in the basic inferring technique), but all the nodes in the \texttt{source set}. Therefore instead of creating a single parent and the flood transaction spending from the same \texttt{utxo}, we will create $n+1$ distinct double spending transactions, $n$ being the number of nodes in the \texttt{source set}: $n$ of those transactions will be tagged as parents, while the remaining one will be the flood transaction. Finally we create a marker transaction from each of the parents, resulting in $n$ parents, $n$ markers, and the flood transaction. Figure~\ref{fig:tx_scheme_all} depicts a high level representation of the created transactions (spending from UTXO$_1$). 

\paragraph{\textbf{Invblock the network:}} Once all the parents, markers and the flood transaction have been created, it is time to ensure that the \textbf{isolation} property will hold during the experiment. It is worth noting that for the isolation property to hold there are two things we need to ensure: First, that the flood transaction ($tx_F$) remains within the \texttt{sink set}. Secondly, that each parent ($tx_{Pi}$) remains only in the \texttt{source set} node ($N_i$) it will be sent to. To ensure so, we will perform an \texttt{invblock} of $tx_F$ and every $tx_P$. \texttt{invblock} consists of sending \texttt{inv} messages to all the nodes in the network with the transaction hashes we want to block the propagation. Recall that a node requesting a transaction with a \texttt{getdata} message in response to a \texttt{inv} message will wait up to two minutes until requesting such transaction to any other peer offering it. By sending multiple \texttt{inv} messages containing the same transaction hashes to a node it can be blocked to request those transactions to any other node for an arbitrary number of minutes, which gives enough time to send all the transactions without having to worry about the isolation property being broken. It is also worth noting that the network will not be blocked with the markers hashes, since their propagation from the \texttt{source set} to the \texttt{sink set} is what will allow us to infer edges between nodes.

\subsection*{Main protocol}
Once we have set the proper conditions for the experiment to be run, we can start sending the transactions we created earlier. 

\paragraph{\textbf{Send transactions:}} First, the flood transaction is sent to all the nodes in the \texttt{sink set}. After waiting a few seconds for the flood transaction to propagate, we can send the proper transactions to the \texttt{source set}. We start by sending a different parent to each node in the set, wait a couple of seconds, and then send the corresponding marker to each node. Since we have \texttt{invblocked} the whole network with the flood and parents, at this point we are sure that, as long as the nodes behave properly, the flood transaction is only present in the \texttt{sink set} and each 
parent is only present in its respective node from the \texttt{source set}.

\paragraph{\textbf{Requesting markers back:}} After waiting a few seconds for the propagation of the markers, we will request all of them back from every node in the \texttt{sink set}. 
%Markers are orphan transactions for nodes in the \texttt{sink set}, so they are stored in the \texttt{MapOrphanTransactions} pool. However, 
Despite being orphans, markers are still considered known transactions by those \texttt{sink set} nodes. In that sense, as we have already seen in the basic inferring technique, when an orphan transaction is requested as part of a \texttt{getdata} message the node holding it will not include it in their response. By sending an \texttt{inv} message containing all the markers to the \texttt{sink set} nodes we will receive back a request of only the subset of markers they have not heard of. \footnote{Notice that some times the subset will be the actual set.} By mapping the markers that have not been sent back to the \texttt{source set} node we originally sent them to, we can infer edges between the \texttt{source set} and the respondent \texttt{sink set} nodes.

\paragraph{\textbf{Permuting the sets:}} With all the aforementioned steps, we are able to infer the edges between a certain configuration of the network, that is, a specific set of nodes forming the \texttt{source set} and \texttt{sink set}. However, the technique cannot infer edges between nodes in the same set. In order to infer the whole topology, we need to run several rounds permuting the sets. Therefore, both the setup and the main protocol will be run until every pair of nodes have been in a different set at least once.

\subsection{Making room for marker transactions.}

The \texttt{MapOrphanTransactions} pool is not allowed to grow unbounded. In
fact, it has a small default limit of only 100 transactions at a time.
When this limit is exceeded, orphan transactions are evicted. The eviction mechanism works as follows: First, it generates a random hash $\mathit{randomhash}$. Next, it selects the transaction in the pool with the closest hash higher than $\mathit{randomhash}$ and evicts it from the pool. The eviction mechanism repeats until the pool size is within limits.

Eviction poses a problem for scaling up the TxProbe technique. Marker transactions must not be evicted until they are read back at the end of a measurement trial.
However, the eviction policy has a design flaw, which enables us to make
preferential transactions that are hard to evict. By crafting transactions for
which their hashes lay between a small fixed range (e.g: by re-signing the transaction), and since the $\mathit{randomhash}$ hash used in the eviction mechanism picks values over an uniform distribution, we can bound the odds of our transaction being evicted depending on how small the range is set.

\paragraph{\textbf{Cleansing the orphan pool:}} 
When we were performing the basic inferring technique there was no need to worry about transactions being evicted from the \texttt{MapOrphanTransactions} pool since we were only creating a single marker. However, now up to $n$ markers would need to be stored by a single node. In order to ensure that there is enough space to store all the markers, we will empty the \texttt{MapOrphanTransactions} pool of all nodes in the \texttt{sink set}. We start by creating a transaction we call the cleanser, and spending from it we create 100 distinct double spending transactions we call the squatters.

Next, we send all the squatters to every single node in the \texttt{sink set}
aiming to full the orphan pool. Finally, we send the cleanser to every
\texttt{sink set} node. Upon reception of the cleanser, all transactions in the
orphan transactions pool will no longer be orphans. One of the squatters will be flagged as valid, whereas the rest will be discarded as double-spending transactions. Regardless of which squatter is accepted by each node, the \texttt{MapOrphanTransactions} pool of each \texttt{sink set node} will be emptied. Figure~\ref{fig:tx_scheme_all} depicts a high level representation of the orphans and cleanser transaction creation (spending from UTXO$_0$). 
%%%%%%%%%%%%%%%%%%%%%%%%%%%%%%%%%%%%%%%%%%%%%%%%%%%%%%%%%%%%%%%%%%%%%%%%%%%%%%%

%%%%%%%%%%%%%%%%%%%%%%%%%%%%%%%%%%%%%%%%%%%%%%%%%%%%%%%%%%%%%%%%%%%%%%%%%%%%%%%
% Costs
%%%%%%%%%%%%%%%%%%%%%%%%%%%%%%%%%%%%%%%%%%%%%%%%%%%%%%%%%%%%%%%%%%%%%%%%%%%%%%%
\section{Costs of topology inferrence}
\label{sec:costs}

In this section we discuss the costs of running TxProbe both in terms of time and transaction fees.

\subsection{Time costs}
How long it takes to infer the topology of a network using TxProbe directly depends on the number of reachable nodes $r_n$ in the network. As we have already seen, the size of our \texttt{source set}s is bound by the \texttt{MapOrphanTransactions} pool size, which is 100 by default. Our set partitioning algorithm works as a grid, in order to separate nodes in two sets we create a grid of width $w = \mathit{min}(\lceil \sqrt{r_n} \, \rceil, 100)$ and length $h = \lceil \dfrac{r_n}{w} \rceil$, and we traverse the grid by rows and columns, being the selected row/column in iteration $i$ our \texttt{source set} for the $i$-th round of the experiment, and the rest of nodes our \texttt{sink set}.\footnote{Notice that when traversing columns the number of elements in the set can be higher than $w$, in which case the algorithm will create $\lceil h/w \rceil$ sets per column.} The total number of different \texttt{source sets}, and therefore, the total number of rounds required to run an experiment will then be:

$$
t_r = 
\begin{cases} 
	h + w - 2,  \mbox{for } h \leq w \\
	h-1+\lceil \dfrac{h}{w} \rceil \cdot w, \mbox{for } h > w  
\end{cases}
$$

\begin{figure}[tb]
	\centering	
	\includegraphics[scale=0.4]{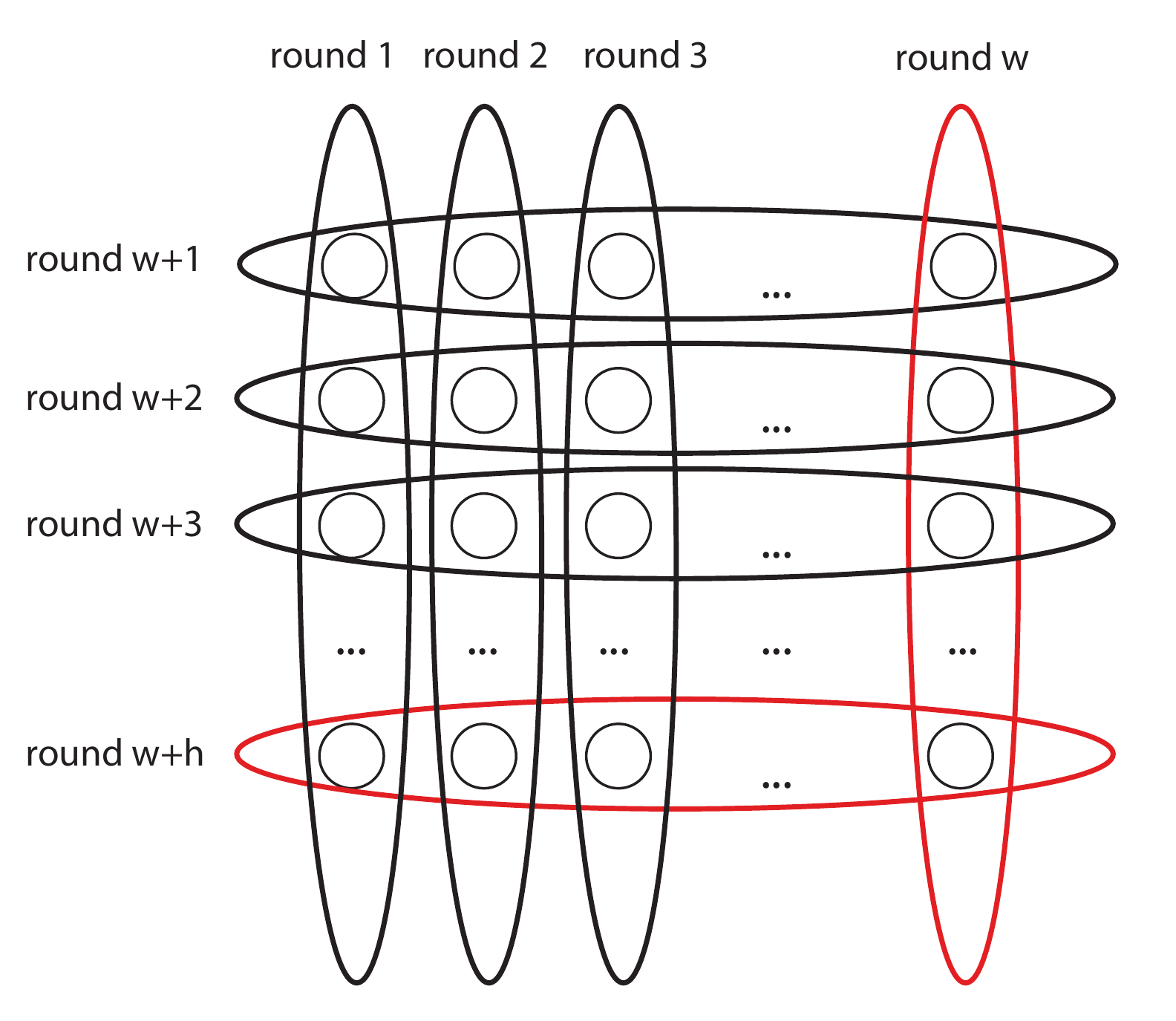}
	\caption{Set-partitioning algorithm used in TxProbe. Sets corresponding to the last row and column (marked in red) can be skipped since they will be already counted by the rest of sets.}
	\label{fig:partitioning}
\end{figure}

Each round of the TxProbe can be run in about 2.5 minutes, resulting in $2.5\cdot t_r$ minutes to run TxProbe over a network of $r_n$ nodes. Inferring the topology of a network like Bitcoin testnet ($\sim1000$ nodes) requires, therefore, about 2.6 hours, whereas inferring the topology of Bitcoin mainnet ($\sim10000$ nodes) requires about 8.25 hours. The partitioning algorithm can be found depicted in Figure \ref{fig:partitioning}.

\subsection{Transaction fee costs}

The costs of running TxProbe directly depend on the number of rounds of the experiment $t_r$ and the fee rate to be paid in order to get our transactions relayed by the network. For every round we will perform an orphan cleansing, resulting in two standard 1-1 P2PKH transactions (only the cleanser and one squatter will be accepted, the rest will be eventually flagged as double-spends). Moreover, at every round either the flood will be accepted (1-1 P2PKH transaction) or a parent-marker pair will be accepted (two 1-1 P2PKH transactions). 

The size of a 1-1 P2PKH transaction using compressed public keys and assuming a signature of maximum length (73-byte signature) is 193 bytes long. Putting all together, the cost of running TxProbe in a network of $r_n$ nodes ranges between $3 \cdot 193 \cdot fee\_rate \cdot t_r$ and $4 \cdot 193 \cdot fee\_rate \cdot t_r$.

At a fee rate of $5 \, sat/byte$ \footnote{Fee to get transactions confirmed between 1-2 blocks on 27th August 2018 according to \url{https://bitcoinfees.earn.com/}.}, running the experiment in a network like Bitcoin mainnet will cost between $573210$ and $764280 \, satoshi$.

\paragraph{Impact of TxProbe measurement.}
We say a few words about the feasibility of using TxProbe to do ethical measurement.
The TxProbe measurement involves sending many kinds of abnormal transactions, and thus it can only be used ethically if we ensure it does not harm or burden the network we are measuring.

To start with, although TxProbe transactions are unusual in that they are multi-way conflicting double-spends, they are not relayed and thus do not increase network or storage utilization compared to ordinary transactions.

The TxProbe experiment can have a destructive effect, however, on the \texttt{MapOrphanTransactions} data structure. As we discussed earlier, if the orphan transactions pool is full, then adding new orphan transactions (such as the marker transactions in TxProbe) can evict others.

We could not rule out the potential that our measurement would add to this congestion (i.e., over an 8+ hours for a scan of the entire network) and could adversely affect real transactions.
%%%%%%%%%%%%%%%%%%%%%%%%%%%%%%%%%%%%%%%%%%%%%%%%%%%%%%%%%%%%%%%%%%%%%%%%%%%%%%%

%%%%%%%%%%%%%%%%%%%%%%%%%%%%%%%%%%%%%%%%%%%%%%%%%%%%%%%%%%%%%%%%%%%%%%%%%%%%%%%
% Experiments and results
%%%%%%%%%%%%%%%%%%%%%%%%%%%%%%%%%%%%%%%%%%%%%%%%%%%%%%%%%%%%%%%%%%%%%%%%%%%%%%%
\section{Experiments and results}
\label{sec:experiments}

We conducted experiments using the Bitcoin testnet in order to evaluate our topology inference technique. We first conducted ground truth experiments to quantify its precision and recall, and then took a snapshot measurement of testnet to demonstrate its usefulness for network-wide scans.

\subsection{Validation}
In order to validate our results we run 5 local Bitcoin nodes as ground truth. The ground truth nodes are included as part of the \texttt{source set} in each round of the experiment. This means that, if our results are correct, at the end of the experiment we would have inferred every edge between the ground truth nodes and the \texttt{sink set} nodes. 

For every run of the experiment there are always nodes that do not behave according to the default client, for example by ignoring \texttt{invblock} and therefore, sending transactions without receiving \texttt{getdata} messages. In order to detect such nodes, an \texttt{invblock} test is performed before every experiment using two Coinscope instances: the first instance crafts a random 32-byte hash and offers it to the whole network using \texttt{inv} messages. The second instance offers the exact same hash within the next two minutes and collects all the \texttt{getdata} responses. All those nodes who responded the second instance are flagged as unblockable nodes and taken out of the experiment. 

Transitory edges (i.e. edges that have been there for a short amount of time) are also removed from the inferred results, as well as nodes who know about transactions they are not supposed to (nodes who missed a parent/marker when it has been sent to them, nodes holding the flood transaction when they were supposed to hold a parent transactions, etc). 	Finally, disconnecting nodes (nodes that have disconnected from Coinscope while the experiment was running) are also removed, as well as all inferred edges referring them.

Our validation reported a precision of $100\%$ and recall between $93.86\%$ and $95.45\%$ with a $95\%$ confidence over 40 runs of the experiment.

\subsection{Analysis of the inferred network}
This section includes a description of a testnet network snapshot taken on 21st February, 2018, as obtained using our technique, which reported a precision of $100\%$ with a recall of $97.40\%$.

% degree

\begin{figure}[tb]
	\centering
	\subfloat[Degree distribution of nodes in the testnet snapshot.\label{fig:deg_dist}]{
	\includegraphics[width=0.5\textwidth]{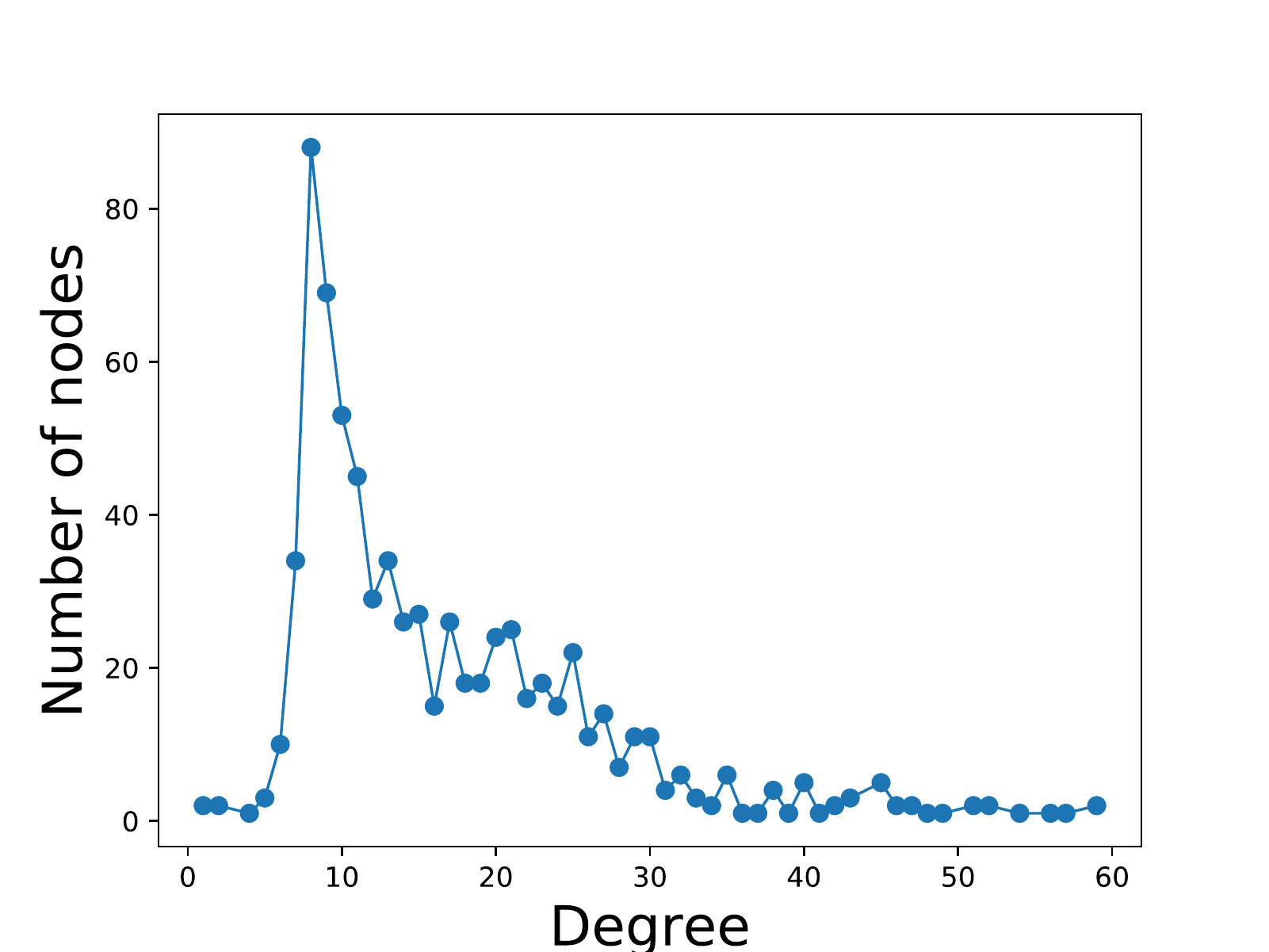}
	}
	\hfill	
	\subfloat[Communities detected in the testnet snapshot.\label{fig:communities}]{
	\includegraphics[width=0.4\textwidth]{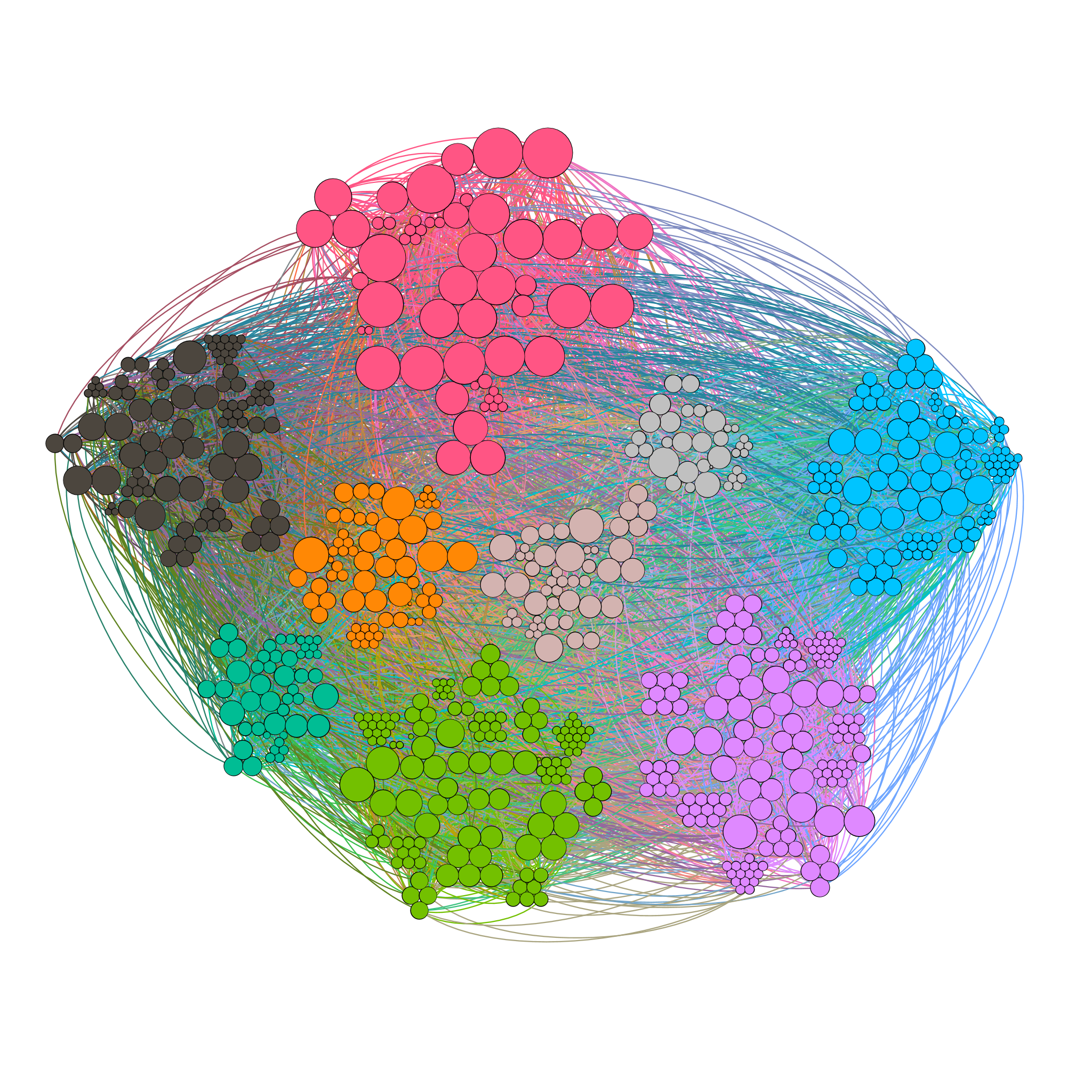}
	}
\end{figure}

The observed network has $733$ nodes and $6090$ edges, with an average degree of $16.6$. The degree distribution of the network is far from uniform (Figure~\ref{fig:deg_dist}), with most of the nodes having between $7$ and $14$ neighbors. The most common degree observed in the network is $8$ (shown by $12\%$ of the nodes), a value that matches the default maximum number of outgoing nodes of the Bitcoin Core client.\footnote{\url{https://github.com/bitcoin/bitcoin/blob/v0.16.2/src/net.h\#L59}} The maximum degree is 59, less than half of the maximum number of default peer connections of Bitcoin Core.\footnote{The maximum number of default connections is set to 125: \url{https://github.com/bitcoin/bitcoin/blob/v0.16.2/src/net.h\#L73}}

% general properties of the graph

Table~\ref{tab:properties} provides a summary of basic properties of the network regarding clustering, distances, assortativity, and community structure, comparing the observed values with those obtained over random graphs with similar characteristics. That is, for each property, we create 100 random graphs that \textit{resemble} the obtained testnet graph, compute the property over the random graphs, and provide both the average of the results over the random graphs and the percentage of times the value over the random graph is higher than the observed in the testnet snapshot. We have considered three different models for generating random graphs: Erd\H{o}s-R\'{e}nyi~\cite{erdos-renyi} (ER), configuration model~\cite{newman2003structure} (CM), and Barab\'{a}si-Albert~\cite{barabasi-albert} (BA). The Erd\H{o}s-R\'{e}nyi model generates graphs where each pair of nodes may have an edge with the same probability, and independently of the other edges of the network. ER generates graphs with a binomial degree distribution, and it is commonly used as baseline to analyse networks. However, the observed testnet graph does not seem to have an ER-like degree distribution (recall Figure~\ref{fig:deg_dist}). Therefore, we also create random graphs using the configuration model, that allows creating networks with a chosen degree distribution. Finally, since many real world computer networks have been reported to be preferential attachment networks, we also include Barab\'{a}si-Albert model. BA generates scale-free networks, that have power-law degree distributions. The random graphs we create \textit{resemble} the observed network: ER graphs are created with the same number of nodes and edges, CM graphs have the same degree distribution (and, therefore, they implicitly also have the same number of nodes and edges), BA graphs have the same number of nodes and a similar number of edges (by adjusting the number of new edges created at each step of the graph generation algorithm, the number of edges is adjusted to be as close as possible to the observed one).

\begin{table}[th]
\setlength{\tabcolsep}{2pt}
\centering
\begin{tabular}[ht]{ c | c | cc  cc  cc }
Metric & Testnet & \multicolumn{2}{c}{ER} & \multicolumn{2}{c}{CM} & \multicolumn{2}{c}{BA}\\
\hline
Diameter 					& $5$ & $4$ & $(0\%)$ & $4.93$ & $(1\%)$ & $4$ & $(0\%)$\\
Periphery size				& $6$ & $612.9$ & $(100\%)$ & $21.2$ & $(80\%)$ & $379.6$ & $(100\%)$\\
Radius 						& $3$ & $3$ & $(0\%)$ & $3$ & $(0\%)$ & $3$ & $(0\%)$\\
Center size					& $45$ & $120.7$ & $(100\%)$ & $57.9$ & $(70\%)$ & $362.9$ & $(100\%)$\\
Eccentricity 				& $3.946$ & $3.827$ & $(0\%)$ & $3.979$ & $(70\%)$ & $3.528$ & $(0\%)$ \\
\hline
Clustering coefficient 		& $0.052$ & $0.023$ & $(0\%)$ & $0.036$ & $(0\%)$ & $0.066$ & $(100\%)$\\
Transitivity 				& $0.128$ & $0.023$ & $(0\%)$ & $0.036$ & $(0\%)$ & $0.057$ & $(0\%)$ \\
\hline
Degree assortativity 		& $0.291$ & $-0.001$ & $(0\%)$ & $-0.008$ & $(0\%)$ & $-0.043$ & $(0\%)$\\
Country assortativity 		& $0.007$ & $-0.001$ & $(0\%)$ & $-0.001$ & $(0\%)$ & $-0.002$ & $(0\%)$ \\
\hline
Clique number 				& $24$ & $3.73$ & $(0\%)$ & $4.05$ & $(0\%)$ & $6.58$ & $(0\%)$\\
Modularity 					& $0.270$ & $0.220$ & $(0\%)$ & $0.216$ & $(0\%)$ & $0.214$ & $(0\%)$\\
\end{tabular}
\caption{Network properties. For random graphs, the 100-run average is provided, with the percentage of times the property over the random graph is higher than the observed in the testnet snapshot in parenthesis.}
\label{tab:properties}
\end{table}

Distances are the properties analyzed in the testnet graph that most approximate those obtained in random graphs. For instance, the radius of the testnet snapshot is 3, exactly the same value observed in all the generated random graphs. The diameter (the maximum distance between any pair of nodes) of the testnet graph is 5, which is higher than most of the random graphs, but very close to their diameters. Moreover, by removing just 3 of the lowest degree nodes of the testnet graph, its diameter becomes 4 (removing the degrees from the sequence in the CM model has the same effect). On the contrary, the number of nodes in the center and in the periphery (i.e. nodes with eccentricity equal to the radius and the diameter, respectively) differs largely from random graphs.

With respect to clustering, the testnet graph exhibits a higher average clustering coefficient than ER and CM graphs, but less than BA graphs. However, observed transitivity is higher than any of the random graphs. Clustering coefficient analyses how well connected the neighborhood of a node is (taking into account the neighborhood regardless of its size), whereas transitivity is focused on studying 3-node substructures.

The testnet snapshot shows higher assortativity than the expected for random graphs, that is, nodes in the testnet tend to connect to other nodes that are similar to themselves more often than what ER, CM and BA random graphs exhibit. Specifically, nodes tend to connect to other nodes with the same degree and, to a less extent but also in a significant manner, to nodes in the same country.

Regarding community structure, we have computed the modularity over the best partition found using the Louvain method. The modularity of the testnet graph is higher than any of the random graphs, regardless of the chosen model. That is, the network shows more community structure than what should be expected for a random graph. Figure~\ref{fig:communities} depicts a visualization of the communities found in the testnet snapshot, with the color of the node denoting the community it belongs to. There are nine communities, with the biggest two (purple and green in the image) having $37\%$ of the nodes of the network. Notably, there is one community (colored in pink) that contains only $7.5\%$ of the nodes but includes the $25$ highest degree nodes of the network. This is consistent with the high degree assortativity reported in the network. Remarkably, the testnet graph contains a clique (a fully connected graph) of $24$ nodes. This clique is found inside the high-degree community (depicted in pink in the visualization). In contrast, the largest clique formed by nodes of any other community has a size of just 6 nodes.

\begin{figure}[tb]
	\centering
	\includegraphics[width=0.8\textwidth]{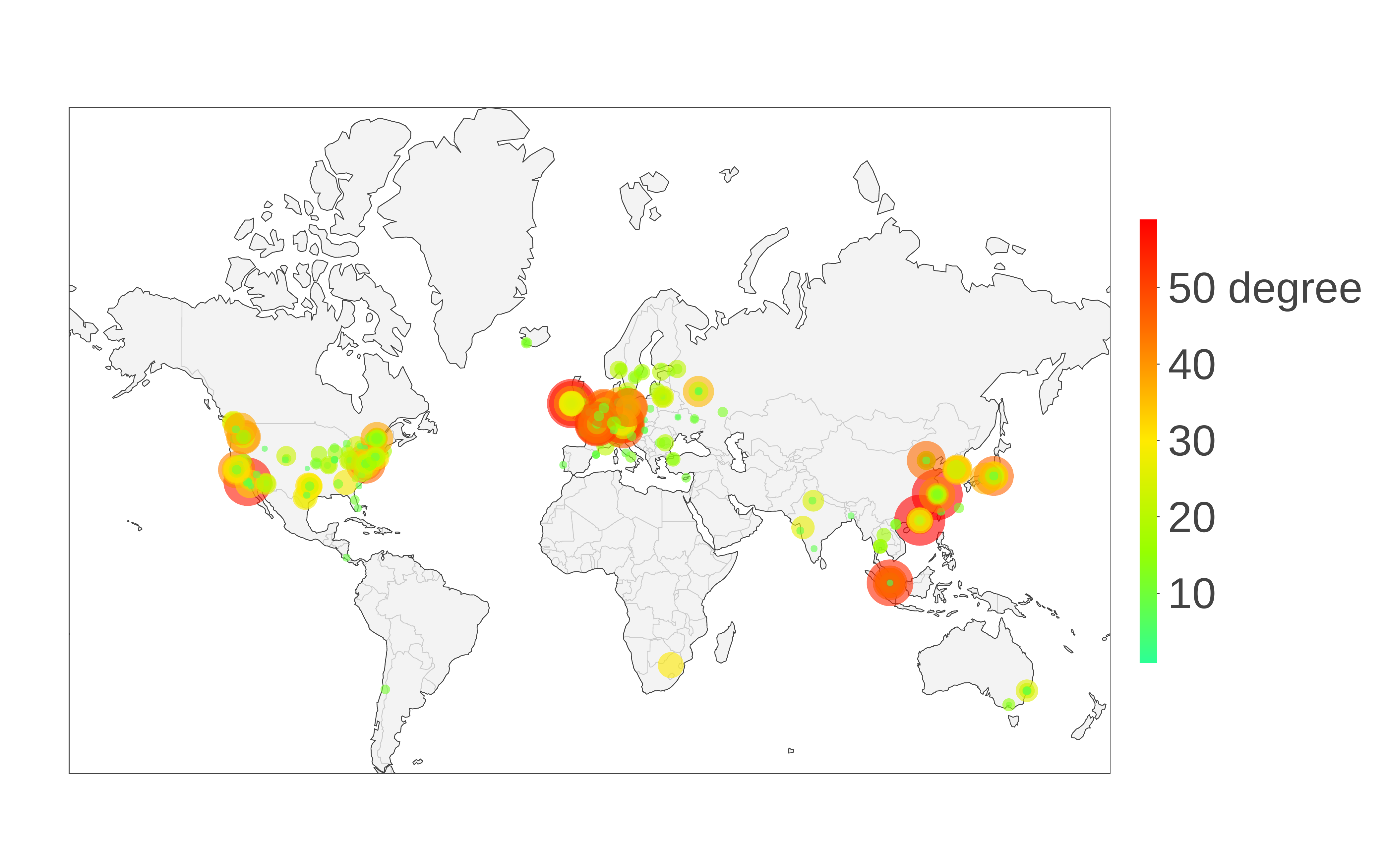}
	\caption{Geographical location of nodes.}
	\label{fig:map}
\end{figure}

We have also used an IP Geolocation API~\footnote{\url{http://ip-api.com/}} to obtain the geographical location of the nodes in the testnet snapshot. Figure~\ref{fig:map} shows a map with the node's locations, where both the size and color of the nodes are used to denote nodes' degree. Most nodes are located in the United States, Central Europe, and East Asia.

%%%%%%%%%%%%%%%%%%%%%%%%%%%%%%%%%%%%%%%%%%%%%%%%%%%%%%%%%%%%%%%%%%%%%%%%%%%%%%%

% Conclusions
%%%%%%%%%%%%%%%%%%%%%%%%%%%%%%%%%%%%%%%%%%%%%%%%%%%%%%%%%%%%%%%%%%%%%%%%%%%%%%%
\section{Conclusions}
\label{sec:conclusions}

We set out to design an effective measurement technique that can reconstruct the Bitcoin network topology. We validated TxProbe to show that it is accurate and can indeed be scaled up to snapshot the entire network  with reasonably low fees.
However, we decided not to carry out a measurement of the main network because we could not rule out its potential to delay the propagation of real users' transactions.
We consider it an open question whether this technique (or analysis thereof) can be improved so it can be used less invasively.

We did, however, take network measurement snapshots of the Bitcoin test network
topology (over 700 nodes). Our analysis of the Bitcoin testnet reveals
significant non-random structure, including several communities, as well as a
clique of high-degree nodes. Although our findings over testnet cannot be
applied to the mainnet, it demonstrates that the technique is viable not only in Bitcoin, but in any other cryptocurrency sharing the network protocol and orphan transaction handling with it.

Like other measurement techniques, TxProbe makes use of implementation-specific behaviors in the Bitcoin software.
While cryptocurrencies have not made topology-hiding an explicit design requirement,
in the past software changes that improve user anonymity have also had the effect of closing
off measurement avenues.
Viewed in this light, TxProbe is the next step in a tacit arms race between measurement efforts and privacy enhancing design.
We make the following suggestions to the cryptocurrency community to avoid this cycle:
First, determine whether network topology or other metrics should be an explicit design goal, in which case effort can be focused on achieving it robustly.
Second, follow the Tor project~\cite{jansen2016safely} for example, in deploying measurement-supporting mechanisms into the software itself, that balances the positive goals of network measurement (such as quantifying decentralization, detecting weaknesses or attacks, etc.) with the privacy goals of users.
%%%%%%%%%%%%%%%%%%%%%%%%%%%%%%%%%%%%%%%%%%%%%%%%%%%%%%%%%%%%%%%%%%%%%%%%%%%%%%%
% Acks
%%%%%%%%%%%%%%%%%%%%%%%%%%%%%%%%%%%%%%%%%%%%%%%%%%%%%%%%%%%%%%%%%%%%%%%%%%%%%%%
\subsubsection*{Acknowledgments}
This work was sponsored in part by a gift from the DTR foundation, and
a grant from by IBM-ILLINOIS Center for Cognitive Computing Systems Research (C3SR)
- a research collaboration as part of the IBM AI Horizons Network.

%%%%%%%%%%%%%%%%%%%%%%%%%%%%%%%%%%%%%%%%%%%%%%%%%%%%%%%%%%%%%%%%%%%%%%%%%%%%%%%

%%%%%%%%%%%%%%%%%%%%%%%%%%%%%%%%%%%%%%%%%%%%%%%%%%%%%%%%%%%%%%%%%%%%%%%%%%%%%%%

\end{document}